\documentclass[showpacs,amsmath,floatfix,prl]{revtex4}
\usepackage{graphicx}
\begin{document}

\begin{center}
\begin{Large}
\textbf{A theoretical and experimental study of the lithiation of $\eta '$-Cu$_6$Sn$_5$ in a lithium-ion battery.}\\
\end{Large}
\bigskip

S. Sharma$^1$, L. Fransson$^3$,  E. Sj\"ostedt$^1$, L. Nordstr\"om$^1$,
B. Johansson$^{1,2}$ and K. Edstr\"om$^3$\\
\textit{
$^1$Condensed Matter Theory Group, Department of Physics, \\
Uppsala University, BOX 530, SE-751 21 Uppsala, Sweden \\
$^2$Applied Materials Physics, \\
Department of Materials Science and Engineering, \\
Royal Institute of Technology, \\ 
SE-100 44 Stockholm, Sweden \\
$^3$Department of Materials Chemistry, \\
Uppsala University, BOX 538, SE-751 21 Uppsala, Sweden \\
}
\end{center}

\centerline{
\begin{minipage}{350pt}
In this work, the mechanism of Li insertion in $\eta '$-Cu$_6$Sn$_5$ to form Li$_2$CuSn is discussed in detail, based on both theoretical 
calculations and experimental results. The mechanism is investigated by means of first principles calculations, with the full potential 
linearized augmented plane wave method, in combination with {\it in situ} X-ray diffraction experiments. The $\eta '$-Cu$_6$Sn$_5$
structure, as well as its lithiated products, have been optimized and the electronic charge density calculated in order to study the 
change in bond character on lithiation. The average insertion voltage of the $\eta '$-Cu$_6$Sn$_5$-Li$_2$CuSn transformation has been 
calculated to be 0.378V in good agreement with the experimental value.
\end{minipage}
}

\newpage

\section{Introduction}

Light weight and  compact, lithium-ion batteries \cite {tarascon01} are ideal energy storage devices for the use in appliances 
like laptops, mobile phones and electric vehicles. But there are concerns about the safety of the lithium-ion batteries in their 
present form because, the most common material used as anodes in these batteries is graphite and its lithiated potential 
is very close to that of the lithium.
The study of intermetallic insertion electrodes, as possible replacements for graphite, has recently attracted a lot of attention. 
Examples of such structures are InSb \cite{johnsson00,vaughey00} and Cu$_2$Sb \cite{fransson01}.
The desirable characteristics for an ideal anode material include the capability of reversible Li intercalation/insertion while
maintaining the stability of the structure.
One of the materials that possess these qualities and has been found to be a possible alternative for graphite in 
lithium-ion batteries is $\eta '$-Cu$_6$Sn$_5$ \cite{thackeray99,kepler99,larcher00,fransson02}. 

{\it In situ} X-ray diffraction (XRD) studies have shown that, during the electrochemical reaction of $\eta '$-Cu$_6$Sn$_5$ with
lithium, a Li$_2$CuSn-type structure is formed \cite{kepler99,larcher00,fransson02}. The ideal reaction for this phase transformation 
can be described by:

\begin{eqnarray}  
{\rm Cu_6Sn_5 + 10Li} \longrightarrow {\rm 5Li_2CuSn + Cu}.
\end{eqnarray}

Further lithiation of the Li$_2$CuSn structure results in further extrusion of Cu from the structure, to finally form Li$_{4.4}$Sn 
(the maximum lithiated Sn-phase). This reaction is described as:

\begin{eqnarray}  
{\rm xLi + Li_2CuSn} \longrightarrow {\rm Li_{2+x}Cu_{1-y}Sn + yCu} \, \, \, \,  (0 \le {\rm x} \le 2.4; 0 \le {\rm y} \le 1).   
\end{eqnarray}

These reactions were recently supported by the {\it in situ} $^{119}$Sn-M\"ossbauer spectroscopy measurements \cite{fransson02}. 
When the lithiation process is interrupted, to only involve reaction (1) (above 0.2V vs. Li/Li$^+$), the material shows a reversible 
and stable lithium insertion behaviour with a capacity of $\sim$ 200mAh/g. If the material is fully lithiated to form Li$_{4.4}$Sn, 
the cycling stability is dramatically affected and the ability to insert lithium reversibly gradually declines. 
The reason for this is believed to originate from the large structural rearrangements that occurs in reaction (2), along with the 
extrusion of Cu. The experimental data provides 
information on the main processes that occurs on lithiation of the $\eta '$-Cu$_6$Sn$_5$ structure. The detailed mechanisms of 
reactions (1) and (2) are, however, unclear. For example the positions of the initial Li in the $\eta '$-Cu$_6$Sn$_5$ structure and 
the starting point of Cu extrusion from the structure are not known. Although there have been suggestions regarding the mechanism 
of the phase change from $\eta '$-Cu$_6$Sn$_5$ to Li$_2$CuSn \cite{thackeray99,kepler99}, there exists no experimental measurement 
or theoretical calculation to back these suggestions. In the present work we have performed {\it ab-initio} calculations to 
investigate this mechanism. The $\eta '$-Cu$_6$Sn$_5$ structure has been optimized and the calculations performed to determine the 
average insertion voltage for the Cu$_6$Sn$_5$-Li$_2$CuSn transformation.

The paper is arranged in the following manner. Section II gives the details of the calculations and experiments. The results of the 
structural optimization and electric field gradients are presented in section III A. Section III B deals with the discussions about the 
charge densities and the nature of the bonds in $\eta '$-Cu$_6$Sn$_5$ and Li$_2$CuSn compounds.  In section III C the results and some 
ideas about the mechanism of lithiation of $\eta '$-Cu$_6$Sn$_5$ are presented. Finally section IV comprises of the conclusions of our
work.  

\section{Methodology}

\subsection{Theoretical}

The total energy calculations are performed using the full potential linear augmented plane wave (FP-LAPW) method using 
the {\sf WIEN97} code \cite{WIEN}. The high-lying Cu and Sn semi-core $d-$states are treated using local orbitals. The calculations 
are performed within the the local density approximation (LDA). The scalar relativistic equations are used in a self-consistent 
scheme.  These calculations are performed using 115 and 204 {\bf k} points in the irreducible Brillouin zone (IBZ) for 
$\eta '$-Cu$_6$Sn$_5$ and Li$_2$CuSn respectively. The structural optimization for $\eta '$-Cu$_6$Sn$_5$ is performed using 4 {\bf k}
points in the IBZ.

$\eta '$-Cu$_6$Sn$_5$ has a monoclinic lattice with the space group $C2/c$ (no.15). The calculations are performed at the 
experimental lattice parameters \cite{larsson94} (a=11.022\AA, b= 7.282\AA, c=9.827\AA ). However, we have used the 
theoretically optimized atom positions given in table I for all the calculations. The conventional unit cell for $\eta '$-Cu$_6$Sn$_5$ 
is presented in figure 1. Structurally Li$_2$CuSn exists in a cubic phase with the space group $F\bar{4}3m$ (no.216) 
\cite{pauly68}. The calculations are performed at the experimental lattice parameter of a=6.282\AA \,  and the atomic positions are 
given in table I.

\subsection{Experimental}

$\eta '$-Cu$_6$Sn$_5$ was synthesised by a high-energy ball-milling procedure, as described elsewhere \cite{kepler99}.  
Electrodes were made by mixing 90 wt\% Cu$_6$Sn$_5$ powder with 5 wt\% carbon black and 5 wt\% EPDM rubber binder.  The slurry was 
spread onto a Ni-foil. 
Two-electrode "coffee-bag" Cu$_6$Sn$_5$/Li-cells were made as described earlier \cite{fransson02}. Cell cycling was performed on a 
Digatron BTS-600 battery tester, in galvanostatic mode, with a current density of 0.022 mA/cm$^2$. 

{\it In situ} X-ray diffraction data of Cu$_6$Sn$_5$ electrodes in Li/Cu$_6$Sn$_5$ cells were collected during the initial discharge 
in transmission mode using a STOE and CIE GmbH STADI powder diffractometer fitted with a position-sensitive detector (CuK$\alpha_1$ 
radiation). Measurements were recorded with a fixed detector covering 38.8$^{\rm o}$ - 45.3$^{\rm o}$ in 2$\theta$ while a constant 
current of 0.15mA was applied by a MacPileII instrument.

%\newpage

\begin{center}

Table I: The optimized and the experimental \cite{larsson94} atomic positions in the $\eta '$-Cu$_6$Sn$_5$ unit cell and the 
experimental atomic positions in the Li$_2$CuSn unit cell \cite{pauly68}. The atomic designations are as in Ref. \onlinecite{larsson94} 
and the site designations are in Wycoff notations \cite{wycoff}.
\medskip

\begin{tabular}{|c|c|c|c|c|}\hline
Compound              & Atom & site  & optimized positions           & experimental positions    \\ 
                      &      &       &  x, y, z                      & x, y, z,                  \\ \hline
$\eta '$-Cu$_6$Sn$_5$ & Cu1  & 8f    &  0.10200, 0.47500, 0.20610    & 0.10096, 0.47297, 0.20236 \\ 
                      & Cu2  & 8f    &  0.30570, 0.50600, 0.60650    & 0.30620, 0.50404, 0.60972 \\ 
                      & Cu3  & 4a    &  0, 0, 0                      & 0, 0, 0                   \\ 
                      & CuA  & 4e    &  0, 0.16800, ${1 \over 4}$    & 0, 0.16020, ${1 \over 4}$ \\ 
                      & Sn1  & 8f    &  0.39000, 0.16500, 0.53000    & 0.39106, 0.16250, 0.52867 \\ 
                      & Sn2  & 8f    &  0.28000, 0.65000, 0.35650    & 0.28518, 0.65499, 0.35792 \\ 
                      & Sn3  & 4e    &  0, 0.80500, ${1 \over 4}$    & 0, 0.79892, ${1 \over 4}$ \\ \hline
Li$_2$CuSn            & Cu   & 4c    &                               & ${1 \over 4}$, ${1 \over 4}$, ${1 \over 4}$ \\
	   	      & Sn   & 4a    &                               & 0, 0, 0                               \\
	  	      & Li   & 24f   &                               & ${1 \over 2}$, 0, 0                    \\
                      & Li   & 4c    &                               & ${3 \over 4}$, ${1 \over 4}$, ${1 \over 4}$ \\ \hline
\end{tabular}

\end{center}

%\newpage

\section {Results and Discussion}

\subsection{The structural optimization and the electric field gradients}

First, we shall discuss the importance of the atomic relaxations in $\eta '$-Cu$_6$Sn$_5$. The base centered monoclinic structure 
of $\eta '$-Cu$_6$Sn$_5$ \cite{larsson94} has four different kinds of Cu atoms at sites 4e, 4a and 8f and three different kinds of Sn 
atoms at site 4e and 8f with a total of 4 formula-units (44 atoms) per conventional unit cell (from now on referred to as the CUC). 
The calculations performed for the experimental atomic positions \cite{larsson94} show that there are forces on 
the atoms which can not be neglected, so the relaxation of the structure is needed. On complete relaxation of the atomic positions, 
there is a gain in energy of 2.19 mRy/formula-unit of Cu$_6$Sn$_5$. The optimized atomic positions are presented in table I. 

The experimentally observed parameter that is sensitive to small structural changes and depends on the electronic charge
distribution in a crystal is the electric field gradient (EFG). We have determined the EFG and the asymmetry parameter for 
both the relaxed and unrelaxed structures for comparison with future experimental work and the values are presented
in table II. The details of the method of calculation of the EFG and the asymmetry parameter using FP-LAPW are given elsewhere
\cite{blaha85,blaha89}. On relaxation of the structure the magnitude of V$_{ZZ}$ for Sn1 and Sn2 increases and decreases for Sn3. 
The asymmetry parameter shows an increase for Sn1 and Sn3 and decrease for Sn2. For all the four kinds of Cu atoms the magnitude of 
V$_{ZZ}$ increases and the asymmetry parameter decreases on the structural relaxation of $\eta '$-Cu$_6$Sn$_5$.

In the case of Li$_2$CuSn there are no forces on the atoms and so the atomic relaxations are not needed and all the further 
calculations are performed with the atoms at the ideal positions. Since the atoms in Li$_2$CuSn are present at high symmetry 
positions of a cubic lattice the theoretical EFG vanishes. 

%\newpage

\begin{center}
Table II: The electric field gradient (V$_{ZZ}$) and the asymmetry parameter ($\eta$) for different Cu and Sn sites in the
$\eta '$-Cu$_6$Sn$_5$ unit cell with the optimized and the experimental \cite{larsson94} atomic positions. 
\medskip

\begin{tabular}{|c|cc|cc|} \hline
                    & Optimized  structure          &          & Experimental structure &  \cite{larsson94}  \\ 
   Atom             & V$_{ZZ}$ (X10$^{21}$ V/m$^2$) & $\eta$   & V$_{ZZ}$ (X10$^{21}$ V/m$^2$) & $\eta$ \\ \hline 
   Cu1              &  -4.05                        & 0.326    &   -3.84                &  0.385       \\ \hline
   Cu2              &  -4.46                        & 0.153    &   -4.32                &  0.163       \\ \hline
   Cu3              &  -4.58                        & 0.257    &   -4.40                &  0.331       \\ \hline
   CuA              &   1.42                        & 0.508    &    1.36                &  0.707       \\ \hline
   Sn1              & -10.35                        & 0.743    &  -10.10                &  0.644       \\ \hline
   Sn2              &   4.18                        & 0.519    &    3.80                &  0.806       \\ \hline
   Sn3              & -14.11                        & 0.718    &  -15.02                &  0.696       \\ \hline 
\end{tabular}
\end{center}

\subsection{The electron charge density}

Figure \ref{dvcdcu6sn5} and \ref{dvcdli2cusn} show the difference between the crystalline charge and the superposed atomic charge 
for $\eta '$-Cu$_6$Sn$_5$ in (102) plane and for Li$_2$CuSn in (110) plane respectively. It is clear from figure 2
 that in case of $\eta '$-Cu$_6$Sn$_5$ the Sn atoms gain a small fraction of the charge lost by the Cu atoms. 
It may be noted that for the Sn atoms  the charge distribution in the inner atomic region (around the core) is spherical, while
the increased charge in the outer atomic region is deformed. 
This deformation of the electron charge density is much more pronounced for the Cu atoms than the Sn atoms. The plot also shows 
the electron cloud between the atoms indicating metallic character of the bonds. 

On the other hand in the case of Li$_2$CuSn, (figure 3 ) the Li atoms are strongly depleted of electrons while the 
Cu atoms show a gain in electronic charge. The nearly spherical shape of the charge distribution indicates an ionic character of 
the bonds.  

A comparison of the figures 2  and 3  indicates the modification in the bonding character. 
Lithiation of $\eta '$-Cu$_6$Sn$_5$ clearly results in a charge transfer from Li to the Cu and Sn atoms to form the more ionic 
Li$_2$CuSn phase. In the battery, Li is inserted as an ion with the electrons supplied from the outer circuit redistributing in the 
structure. In this paper we have chosen to use the term Li atoms, since the calculations are based on Li atoms within the structures.

\subsection {The lithiation of $\eta '$-Cu$_6$Sn$_5$} 

According to the reactions (1) and (2), the total number of Li atoms reacting with one formula unit of Cu$_6$Sn$_5$ is 22. The 
experimental voltage {\it vs.} composition curve (figure 4) shows a maximum x of almost 22, which is in very good agreement 
with the suggested mechanisms. The {\it in situ}-XRD and M\"ossbauer measurements have, however, shown that the reactions taking 
place above 0.7V (section A in figure 4) can be attributed to surface reactions and reduction of oxides present in the sample 
due to the synthesis method (high-energy ball-milling) \cite{fransson02}. These reactions are irreversibly consuming Li and are not 
directly involved in reactions (1) and (2). The total experimental value for these reactions is thereby slightly smaller. The plateau, 
at around 0.4V (in section C of figure 4), is associated with the transformation of $\eta '$-Cu$_6$Sn$_5$ to Li$_2$CuSn.
It has been suggested, that during this phase transformation, half of the Sn atoms in the $\eta '$-Cu$_6$Sn$_5$ structure are displaced 
into neighbouring Sn-strings \cite{thackeray99,kepler99}. 
The Li$_2$CuSn-type phase at the end of the plateau is most likely lithium deficient, as further lithiation down to 0.2V 
(the sloping part in the voltage curve) results in a gradual expansion of the Li$_2$CuSn cubic cell axis, {\it i.e.} a solid-solution 
behaviour within the structure. The process below 0.2V (section D in figure 4) involves lithiation of Sn to finally yield 
Li$_{4.4}$Sn. We have focused mainly on the process described by reaction (1) (the Cu$_6$Sn$_5$- Li$_2$CuSn transformation). 

From the total energy calculations the average insertion voltage (AIV) for the Cu$_6$Sn$_5$-Li$_2$CuSn transformation is calculated.  
The details of the method of calculation of AIV are given elsewhere 
\cite{ceder97,deiss99}. The calculated value of the AIV is 0.378V which, is in good agreement with the experimental value of the plateau in 
the voltage profile (figure 4 ).

In the voltage profile, the region between 0.7V and 0.4V (section B in the figure 4), {\it  i.e.} after the reduction of 
the oxides present in the sample and before the start of the plateau, there is a sloping region involving approximately two Li atoms. 
This region would then correspond to the first two Li atoms inserted in the $\eta '$-Cu$_6$Sn$_5$ structure ({\it i.e.} x=1 to x=2 in 
the reaction (1)).
The {\it in situ} XRD measurements in this region show that the first lithium insertion does not result in any major structural 
rearrangements and the $\eta '$-Cu$_6$Sn$_5$ peaks remain unaltered. The first-principles total energy and force calculations have been 
performed to study the various possibilities of how these two Li atoms 
are inserted into the $\eta '$-Cu$_6$Sn$_5$ structure. In figure 5 the calculated total valence charge density of 
$\eta '$-Cu$_6$Sn$_5$ in the (102) plane is plotted (there exist four such planes per CUC).
The minimum densities are found in the regions marked by A and B in the figure (there are 8 crystallographically symmetric sites of 
each kind per CUC). By looking at the charge density plots in various planes in the unit cell (not presented in the paper to 
keep the number of figures limited) the regions A, B and C are found to have minimum electron density and hence represents voids 
where the Li atoms can enter. 

With this information we go on to study how well these sites incorporates a Li atom. By total energy calculations it was 
found that the energy requirement for the insertion of a Li atom at site A in the CUC (one Li atom per CUC corresponds to x=0.25 
in reaction (1) and x=3 in figure 4) is 41mRy/Li-atom. On the other hand the energy required for insertion of a Li atom 
at site B or C in the CUC is 66mRy/Li-atom and 120mRy/Li-atom respectively. This shows that it is energetically favourable for the 
first two Li atoms to enter at the sites A rather than the sites B or C. 

On complete relaxation of the structure with eight Li atoms per CUC (making a total of 52 atoms per CUC, {\it  i.e.} x=2.0 in reaction
(1) and x=4.5 or end of section B in figure 4) at the eight crystallographically symmetric A sites, two of which are marked
 in figure 5, the gain in energy is 190mRy per Li$_2$Cu$_6$Sn$_5$ formula-unit with a volume expansion of 4.6\%. 
In this relaxation the Sn (Sn2) atoms move towards the site B, occupying the position with coordinates: 0.643, 0.450, 0.200 and the  
Li atoms move to the position earlier occupied by the Sn (Sn2) atoms. This picture supports the earlier suggestions 
\cite{thackeray99,kepler99} 
about the structural rearrangement on lithiation of $\eta '$-Cu$_6$Sn$_5$. However, with just two Li atoms per formula-unit there is
no major structural rearrangement and the Sn atoms do not reach the sites B to form the Sn chains, but move towards sites B and come to 
rest nearly half way between B and its original position (Sn2). This rearrangement is not detected experimentally. 

There are two scenarios for further lithiation, one where lithium is inserted without any copper extrusion and the other where
the lithium insertion is accompanied by extrusion of the CuA atoms (figure 5) from the structure. There exists no
information about the morphology and the phase adopted by the extruded copper atoms, so the second case can not be 
studied theoretically. It was found that the first case (lithium insertion without copper extrusion {\it i.e.} 3 Li atoms per 
Cu$_6$Sn$_5$ formula-unit) is accompanied by a volume expansion of 47.3 \% and costs large amount of energy (290mRy/Li-atom) which in 
turn would imply a sloping voltage profile in the beginning of the section C {\it i.e.} between x=4.5 to x=5.5 in the figure 4 
(which corresponds to x=2 to x=3 in reaction (1)). However, experimental voltage profile instead shows a  plateau after insertion of 2 Li 
atoms per Cu$_6$Sn$_5$ formula-unit. 
We, therefore, suggest that after the insertion of two Li atoms per Cu$_6$Sn$_5$ formula unit, the Cu (CuA) atoms start to extrude from 
the structure accompanied by structural rearrangements.

{\it In situ} XRD measurements were performed to closely follow the phase transformation from $\eta '$-Cu$_6$Sn$_5$ to Li$_2$CuSn. 
The results can be seen in figure 6, showing diffraction patterns taken at consecutive steps at the 0.4V plateau ({\it i.e.}
beyond 2 Li atoms per formula-unit). On lithiation, the Cu$_6$Sn$_5$ peak at $\sim$ 43$^{\rm o}$ gradually decrease in intensity, while 
the Li$_2$CuSn peak at $\sim$ 40.5$^{\rm o}$ starts to increase in intensity. The data shows that the plateau corresponds to the two-phase 
transformation between Cu$_6$Sn$_5$ and Li$_2$CuSn, where half of the Sn atoms move to the neighbouring Sn chains. This transformation is 
accompanied by quite large structural rearrangements. The Cu$_6$Sn$_5$ peak decreases in intensity and broadens before the first sign of 
the Li$_2$CuSn phase. The diffraction peak of the resulting Li$_2$CuSn phase is even broader due to a more disordered phase. This supports 
the theoretical picture presented above.

The M\"ossbauer and XRD \cite{fransson02,erik02} experiments, performed for phase determination during 
lithiation of $\eta '$-Cu$_6$Sn$_5$, are not sensitive to the point of copper extrusion and the phase adopted by the 
extruded copper atoms. Since the cycling stability depends upon the copper extrusion and stability of the structure 
we suggest EXAFS (Extended X-ray Absorption Fine-structure Spectroscopy) and HRTEM (High-Resolution Transmission Electron Microscopy) 
be performed to provide more information on the exact starting point of the copper extrusion as well as the structure and 
morphology of the extruded copper. Further, the knowledge of the phase of the extruded copper atoms would facilitate future theoretical 
work on the copper deficient lithiated $\eta '$-Cu$_6$Sn$_5$.

\section{Conclusions}

$\eta '$-Cu$_6$Sn$_5$ is a candidate anode material for lithium-ion batteries. At a discharge voltage above 0.2V lithium-ions are inserted 
in $\eta '$-Cu$_6$Sn$_5$ reversibly to form Li$_2$CuSn. In this region the cycling capacity is stable around 200 mAh/g. 

From the first principles total energy and force calculations the atomic positions in $\eta '$-Cu$_6$Sn$_5$ are optimised and EFG 
(the structure and electronic charge dependent parameter) is determined. The AIV of the $\eta '$-Cu$_6$Sn$_5$ to Li$_2$CuSn 
transformation has been calculated to be 0.378V which is in good agreement with the experimental result of $\sim$ 0.4V. 

By calculating and plotting the difference between the valence- and atomic charge in a crystal plane, the change in the bonding nature 
on lithiation of $\eta '$-Cu$_6$Sn$_5$ could be studied. It was found that the ionicity of the bonds increases on going from 
$\eta '$-Cu$_6$Sn$_5$ (where the bonds have metallic character) to Li$_2$CuSn.

Total energy and force calculations show that it is energetically favourable for the first two incoming Li atoms (two Li atoms per 
$\eta '$-Cu$_6$Sn$_5$ formula-unit) to enter at the crystallographically symmetric sites with the positional coordinate 
0.907, 0.625, 0.0408. Complete relaxation of the structure leads to small structural rearrangement with the relaxation of the Sn (Sn2) 
atoms to occupy sites with the positional coordinate 0.643, 0.450, 0.200 and Li atoms to the positions earlier occupied by the Sn (Sn2)
atoms. This leads to a large gain in energy of 190mRy/formula-unit of Li$_2$Cu$_6$Sn$_5$ with a small volume expansion of 4.6\%.  
Further addition of Li atoms (after two Li atoms per Cu$_6$Sn$_5$ formula-unit) without any copper extrusion from the structure costs a 
large amount of energy leading to a sloping voltage profile. However, the experimental voltage profile shows a plateau after x=4.5 
(which corresponds to x=2.0 in reaction (1)), which suggests that after the insertion of two Li atoms per Cu$_6$Sn$_5$ formula-unit 
(eight Li atoms per CUC) the copper atoms are extruded form the structure leading to large structural rearrangement. This picture is 
supported by our XRD measurements which does not show any detectable structural change up to 2 Li atoms per Cu$_6$Sn$_5$  formula-unit
but indicates a major structural rearrangement beyond that.

\section {Acknowledgments} 
Support for this work from The Foundation for Environmental Strategic Research (MISTRA), The Nordic Research Programme (NERP) and 
The Swedish Science Council (VR) is gratefully acknowledged. We are also grateful to the Swedish National Supercomputer Center in 
Link\"oping (NSC) for the use of their computational facilities. We also thank Dr. M. M. Thackeray and Dr. R. Benedek from Argonne 
National Laboratory, USA, for valuable comments and suggestions.

%\newpage
\section{Figure captions}

\begin{itemize}
\item Fig 1. The conventional unit cell of $\eta '$-Cu$_6$Sn$_5$. The (102) plane is marked. 

\item Fig 2. The difference between the crystalline valence charge and the superposed atomic charge for $\eta '$-Cu$_6$Sn$_5$
in the (102) plane. The difference is given in the units of charge per $a_0^3$, with the negative numbers 
in the legend indicating an increase in electrons.

\item Fig 3. The difference between the crystalline valence charge and the superposed atomic charge in the (110) plane of the
Li$_2$CuSn unit cell. The difference is given in the units of charge per $a_0^3$, with the negative numbers in the legend 
indicating an increase in electrons.

\item Fig 4. The voltage profile for the first discharge of a Li/Cu$_6$Sn$_5$ cell, x indicates number of Li atoms inserted per 
Cu$_6$Sn$_5$ formula unit.

\item Fig 5. The total valence electron charge density per $a_0^3$ for $\eta '$-Cu$_6$Sn$_5$ in the (102) plane.
The six regions of low charge density are marked as A, B and C.

\item Fig 6. The {\it in situ} XRD data of a Li/Cu$_6$Sn$_5$ cell measured at the 0.4V plateau during 
galvanostatic discharge. The Ni peak shown is from the current collector.

\end{itemize}


\begin{thebibliography}{10}

\bibitem{tarascon01}
J-M. Tarascon and M.~Armand.
\newblock {\em Nature}, 414:359, 2001.

\bibitem{johnsson00}
C.~S. Johnsson, J.~T. Vaughey, M.~M. Thackeray, T.~Sarakonsri, S.~A. Hackney,
  L.~Fransson, K.~Edstr{\"o}m, and J.~O. Thomas.
\newblock {\em Electrochem. Comm.}, 2:595, 2000.

\bibitem{vaughey00}
J.~T. Vaughey, J.~O'Hara, and M.~M. Thackeray.
\newblock {\em Electrochem. Solid State Lett.}, 3:13, 2000.

\bibitem{fransson01}
L.~M.~L. Fransson, J.~T. Vaughey, R.~Benedek, K.~Edstr{\"o}m, J.~O. Thomas, and
  M.~M. Thackeray.
\newblock {\em Electrochem. Comm.}, 3:317, 2001.

\bibitem{thackeray99}
M.~M. Thackeray, J.~T. Vaughey, A.~J. Kahaian, K.~D. Kepler, and R.~Benedek.
\newblock {\em Electrochem. Commun.}, 1:111, 1999.

\bibitem{kepler99}
K.~D. Kepler, J.~T. Vaughey, and M.~M. Thackeray.
\newblock {\em Electrochem. Solid State Lett.}, 2:307, 1999.

\bibitem{larcher00}
D.~Larcher, L.~Y. Beaulieu, D.~D. MacNeil, and J.~R. Dahn.
\newblock {\em J. Electrochem. Soc.}, 147:1658, 2000.

\bibitem{fransson02}
L. Fransson, E. Nordstr{\"o}m, K. Edstr{\"o}m, L. H{\"a}ggstr{\"o}m and M. M.
  Thackeray accepted for publication in {\it J. Electrochem. Soc.} 2002.

\bibitem{WIEN}
P. Blaha, K. Schwarz and J. Luitz, WIEN97, Vienna university of technology
  (improved and updated UNIX version of the orignal copyrighted wien code),
  published by P. Blaha, K. Schwarz, p. Sorantin, and S. B. Trickey, Comput.
  Phys. Commun. {\bf 59} (1990) 399.

\bibitem{larsson94}
A.~K. Larsson, L.~Stenberg, and S.~Lidin.
\newblock {\em Acta. Cryst. B}, 50:636, 1994.

\bibitem{pauly68}
H.~Pauly, A.~Weiss, and H.~Witte.
\newblock {\em Z. Metallkde.}, 59:47, 1968.

\bibitem{wycoff}
R. W. G. Wyckoff in, 2nd ed. Crystal Structures, Vol 1, Krieger, FL, (1982).

\bibitem{blaha85}
P.~Blaha, K.~Schwarz, and P.~Herzig.
\newblock {\em Phys. Rev. Lett.}, 54:1192, 1985.

\bibitem{blaha89}
P.~Blaha, P.~Sorantin, C.~Ambrosch, and K.~Schwarz.
\newblock {\em Hyp. Int.}, 51:917, 1989.

\bibitem{ceder97}
G.~Ceder, M.~K. Aydinol, and A.~F. Kohan.
\newblock {\em Comput. Mater. Sci.}, 8:161, 1997.

\bibitem{deiss99}
L.~Benco, J.~L. Barras, M.~Atanasov, C.~Daul, and E.~Deiss.
\newblock {\em J. Solid State Chem.}, 145:503, 1999.

\bibitem{erik02}
E. Nordstr{\"o}m, S. Sharma, E. Sj{\"o}stedt, L. Fransson, L.
  H{\"a}ggstr{\"o}m, L. Nordstr{\"o}m and K. Edstr{\"o}m accepted for
  publication in {\it Hyperfine Interact.} 2002.

\end{thebibliography}
\end{document}